\documentclass[conference,compsoc]{IEEEtran}
\ifCLASSOPTIONcompsoc
  \usepackage[nocompress]{cite}
\else
  \usepackage{cite}
\fi
\ifCLASSINFOpdf
\else
\fi
\usepackage{caption}
\usepackage{subcaption}
\usepackage{multirow}
\usepackage{enumitem}
\usepackage{hyperref}
\usepackage{xcolor}
\usepackage{rotating}
\usepackage{algorithm}
\usepackage{algpseudocode}
\usepackage{tabularx}
\usepackage{array}
\usepackage{float}
\usepackage{booktabs}
\usepackage{algorithm}
\usepackage{algpseudocode}
\usepackage{amsmath}
\usepackage{supertabular}
\usepackage{longtable}
\usepackage[table]{xcolor}
\usepackage{xurl}
\usepackage{adjustbox}

\definecolor{ledgernavy}{HTML}{173D67}
\definecolor{ledgerblue}{HTML}{2465A0}
\definecolor{numberblue}{RGB}{20,110,220}
\definecolor{purpletext}{RGB}{128,0,128}
\definecolor{lightgray}{HTML}{F0F0F5}

\definecolor{reportblue}{HTML}{194D86}
\definecolor{reportlight}{HTML}{F7FAFE}
\definecolor{reportpurple}{HTML}{91356C}

\usepackage[most]{tcolorbox}
\usepackage{listings}

\newif\ifcomment
\commentfalse

\ifcomment
\newcommand{\jaishnoor}[1]{{\bf \textcolor{blue}{Jaishnoor: #1}}}
\newcommand{\sayak}[1]{{\bf \textcolor{red}{Sayak: #1}}}

\else
\newcommand{\jaishnoor}[1]{}
\newcommand{\sayak}[1]{}
\newcommand{\purple}[1]{}
\fi

\begin{document}
%
\title{ReDence: A Hybrid Agentic and Decision-Model Framework for Rapid, Evidence-Grounded Detection of Modern Social Engineering Scams}



%

\author{
{\rm Jaishnoor Kaur, Josef Horáček, Ian Waskom, Sayak Saha Roy}\\
 Louisiana State University \\
{\rm\{Jaishnoor.Kaur, jhoracek, Ian.Waskom\}@lsu.edu, sayak.saharoy@lsu.edu}
}


\maketitle

\begin{abstract}
Detecting online social engineering (SE) scams often requires examining website workflows, verifying claims against external sources, and connecting evidence across multiple pages. Existing detectors provide limited support for these investigations, leaving analysts to manually gather and validate evidence, which inevitably delays detection and subsequent takedown of these attacks.

To address this gap, we present ReDence, an open-source hybrid framework for adaptive, evidence-grounded detection across diverse SE scam categories. ReDence combines two locally hosted models with complementary responsibilities: a fine-tuned multimodal agentic Investigator (Qwen3.5-4B) that performs investigative planning, evidence interpretation, and reporting, and a specialized Decider (Tev1) that handles binary and multiple-choice tasks, including browser-action selection, evidence assessment, and decisions about whether to continue an investigation.
By delegating discrete decisions to a specialized model, ReDence reduces reliance on lengthy language-model responses, while preserving the reasoning capabilities required for complex investigations. 

Our evaluation demonstrates that ReDence achieves 96\% accuracy in detecting scams across twelve SE scam categories, substantially outperforming existing detectors while being at least four times faster than comparable agentic models without compromising investigative depth. 

ReDence also produces structured, evidence-grounded reports that document investigative actions, supporting and contradictory evidence, and collected artifacts, which allow analysts to independently audit detection decisions and utilize the collected evidence to support subsequent mitigation and takedown efforts.

\end{abstract}
\section{Introduction}

Social engineering (SE) scams remain a major source of internet-enabled financial harm. Although prior research has extensively characterized different SE scam categories \cite{bitaab2023beyond, onwubiko2020fraud, stabek2010seven, escayola2025typology}, automated website-level defenses have largely focused on \emph{credential-phishing attacks} \cite{liu2023knowledge, abdelnabi2020visualphishnet, roy2025explain}, which often exhibit recognizable indicators of impersonation or credential harvesting on their landing pages~\cite{liu2022inferring,lin2021phishpedia, roy2025explain},

Yet credential phishing represents only one component of a broader SE scam landscape. The FBI's 2025 Internet Crime Report documents \$215.8 million in losses attributed to phishing, compared with \$503.4 million from non-payment/non-delivery scams, \$8.6 Billion from investment fraud, and \$275.1 million from real-estate fraud~\cite{ic3-2025}. 

However, detecting these SE scams often requires navigating multiple pages, interacting with website workflows, examining supporting artifacts, and independently verifying claims against external sources. For example, \textsf{usastudentdebtrelief.com} portrayed itself as a legitimate student-loan relief service while allegedly collecting more than \$7.3 million in unlawful advance fees and payments for services never rendered~\cite{ftc2025studentloan}. According to the FTC, the operation falsely claimed affiliation with the U.S. Department of Education, advertised repayment terms inconsistent with federal programs, and fabricated customer testimonials on both its own website and third-party platforms~\cite{ftc2024usasdr}. Importantly, evidence contradicting these claims could have been uncovered by cross-referencing information from multiple independent sources, even during the attack's \textit{zero-day} period. Yet, these warning signs were overlooked, illustrating how scams can evade detection when their deceptive claims are not systematically investigated and independently verified.

Recent systems have begun extending automated detection beyond credential phishing, including approaches targeting particular scam categories~\cite{bitaab2025scammagnifier,li2023double} and systems for discovering candidate scam websites~\cite{paudel2025loki}. However, detectors that primarily return classification labels or confidence scores provide limited insight into the evidence underlying their assessments \cite{bitaab2025scamnet}. Such outputs may be insufficient for validating suspected threats or supporting downstream interventions~\cite{malach2025cybershapley,oest2020phishtime}. Consequently, analysts must often manually reconstruct the attack workflow and any necessary evidence before initiating reporting or takedown procedures. This process is time-consuming, difficult to scale, and dependent on familiarity with particular scam schemes or regional contexts \cite{moritz2025operationalizing }, inevitably allowing malicious websites to remain active while, ideally, the investigations are completed, or the attack is misdetected to be benign.

The advent of agentic large language models (LLMs) offers a promising means of automating such investigations in depth~\cite{nakano2025scamferret}, but introduces two challenges. \emph{First}, the ability to reason and invoke tools does not inherently ensure effective end-to-end security investigation~\cite{deng2024pentestgpt}. General-purpose models may require domain-specific investigative guidance to apply relevant knowledge, recognize deceptive practices, pursue appropriate verification steps, and distinguish suspicious observations from substantiated findings~\cite{ye2025sop,nakano2025scamferret}. 

\emph{Second}, comprehensive agentic investigations are computationally expensive, as they involve repeated decisions about tool selection, evidence interpretation, and investigative continuation, often requiring multiple generative LLM invocations even for simple binary or multiple-choice tasks \cite{zhang2026llms, ghoshal2026tools}. The resulting latency, computational overhead, and API costs significantly limit their practicality for large-scale scam detection \cite{ghoshal2026tools, trad2026toward}.

We address these challenges by introducing \textbf{ReDence}, a highly scalable hybrid framework designed to adaptively investigate and detect SE scams. ReDence integrates a fine-tuned multimodal agentic model (based on Qwen3.5-4B~\cite{qwen3.5}), responsible for investigative planning, reasoning, and evidence interpretation, with a specialized Decision model (based on Tev1~\cite{tev1}) that rapidly (and efficiently) handles  binary and multiple-choice decisions throughout the investigation, which significantly reduces unncessary generative overheads. These models are also equipped with specialized investigative prompts and tools that enable systematic, purpose-driven exploration of websites and their associated internal and external artifacts, while dynamically adapting investigative strategies based on emerging evidence and findings.

ReDence also maintains an \emph{evidence-grounded record} of its investigation through reports that document the actions performed, artifacts collected, supporting and contradictory evidence, external verification results, and the reasoning connecting these observations to its final assessment. This design allows a human analyst to inspect not only \emph{what} ReDence concluded, but also \emph{how} it reached that conclusion and the evidence on which it relied. 
In doing so, our framework aims to combine the scalability of automated detection with the adaptive investigation and evidentiary record traditionally provided through human analysis.

We evaluate ReDence on SE scams spanning twelve categories, achieving \textbf{96\% overall detection accuracy}, outperforming the evaluated automated detectors while completing investigations at least four times faster than comparable agentic configurations. We further assess the quality of ReDence's investigative reports through human evaluation.

The primary contributions of this work are as follows:

\begin{enumerate}
    \item We introduce \textbf{ReDence}, an open-source hybrid framework that integrates a fine-tuned multimodal agentic model with a specialized Decision model, supported by domain-specific investigative prompts and tools. This design enables adaptive, evidence-grounded investigations of SE scams while substantially reducing computational overhead.

    \item We develop a structured reporting mechanism that allows ReDence to document each investigation through detailed reports and an evidence ledger, linking investigative findings and assessments to observations, external sources, collected artifacts, and supporting or contradictory evidence. We further assess the relevance and auditability of these reports through human evaluation, finding them to not only enable analysts to identify new scams, but also identify more evidence points in scams they had previously detected.

    \item We comprehensively evaluate ReDence against existing automated scam detectors across twelve SE scam categories. ReDence achieves \textbf{96\% detection accuracy}, substantially outperforming existing approaches while delivering at least a \textbf{fourfold speedup} over comparable agentic configurations without compromising investigative depth.
\end{enumerate}

\section{Background}

\textbf{Automated scam detectors:} 
Despite the prevalence and diversity of SE scams, relatively few automated detection systems have been developed to address them. Existing approaches also vary considerably in their coverage of scam categories and ability to conduct in-depth investigations. For example, BeyondPhish~\cite{bitaab2023beyond} detects fraudulent websites using predefined features extracted from website content, URLs, domain information, and associated social-media accounts, but provides classification decisions without supporting investigative reasoning. ScamMagnifier~\cite{bitaab2025scammagnifier} enables deeper investigations by navigating checkout workflows and examining merchant and payment infrastructure, although its capabilities are specifically tailored to fraudulent shopping websites. Meanwhile, LOKI~\cite{paudel2025loki} focuses primarily on discovering scam websites through learned search queries rather than investigating individual websites.

ScamFerret~\cite{nakano2025scamferret} is the closest prior work to ReDence. It leverages commercial LLMs to detect SE scams by analyzing textual website artifacts and invoking a predefined set of investigative tools. However, ScamFerret cannot directly interact with website workflows or associated artifacts, and its investigative approach does not consistently adapt to category-specific indicators of deception. In Section~\ref{sec:evaluation}, we compare ReDence against ScamFerret and demonstrate how these limitations hinder its ability to systematically uncover and verify evidence necessary for detecting diverse SE scams.

\textbf{Decision models:} General-purpose LLMs often generate lengthy reasoning traces and explanations, even for tasks that require only a binary answer or a selection among predefined alternatives. When repeated throughout an agentic workflow, this unnecessary token generation can introduce substantial latency and computational overhead \cite{madaan2025rethinking}.

To address this inefficiency, recent work has introduced specialized \emph{Decision models}, designed to efficiently handle tasks with a bounded set of possible outcomes~\cite{together2026tev1,ollama2026decision}. Given contextual information, a question, and a set of candidate answers, these models directly score the available alternatives and return a compact decision rather than generating an extended textual response. This makes them particularly suitable for recurring tasks such as action selection, request routing, and content assessment. Recent evaluations highlight their potential for low-latency decision-making. Deußer et al.~\cite{deusser2026evaluating} report a mean client-side latency of 0.36 seconds per request, including network overhead, when evaluating Jev across 37 datasets with 32 concurrent requests. Similarly, Ollama reports that Nimble-9B achieved an average latency of 91 milliseconds per decision in a game-playing demonstration conducted on an M5 Max~\cite{ollama2026decision}. 

ReDence applies this specialization through its hybrid architecture, in which the \emph{Investigator} delegates tasks requiring discrete decisions or selections among predefined alternatives to the specialized \emph{Decider} while retaining responsibility for developing investigative plans, interpreting relationships among findings, and generating evidence-grounded reports. This division combines rapid decision-making with the sustained reasoning and adaptability required for comprehensively evaluating SE scams.
\newline

\section{Research overview}

\begin{figure*}[t]
    \centering
    \includegraphics[width=\textwidth]{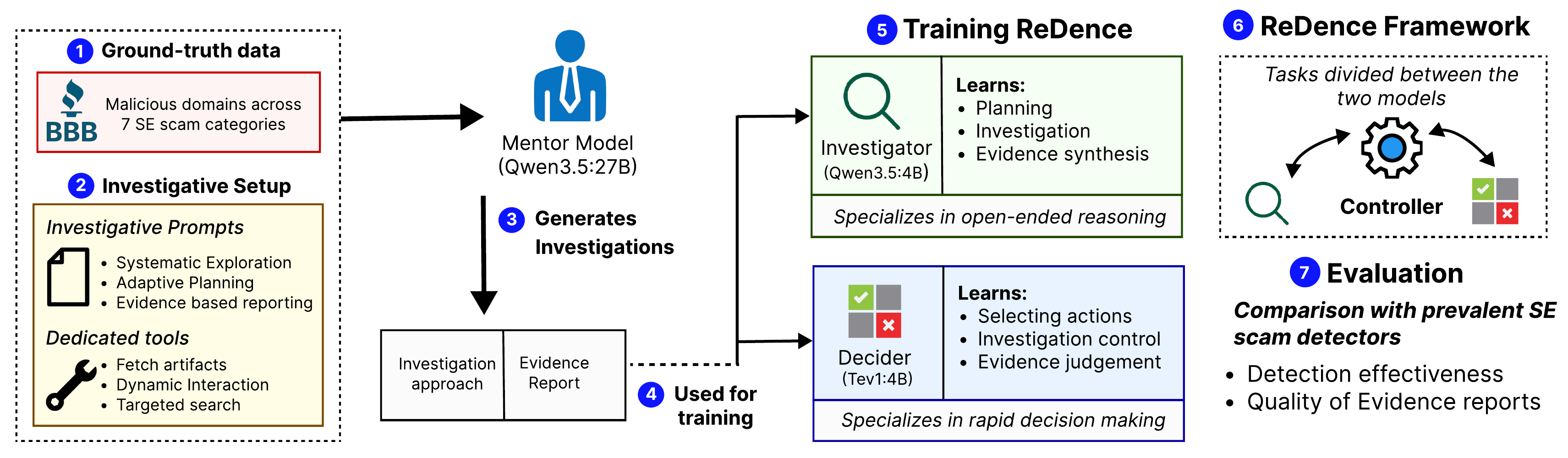}
    \caption{Our framework for developing and evaluating ReDence}
    \label{fig:framework}
\end{figure*}

ReDence combines two compact models, each with 4 billion parameters: a Qwen3.5-based \emph{Investigator} for open-ended reasoning and reporting, and a Tev1-based \emph{Decider} for binary and multiple-choice decisions. Although this architecture offers substantial advantages in computational efficiency and inference speed, prior research has documented limitations in smaller models' ability to follow complex instructions~\cite{jiang2024followbench,wen2407benchmarking}, coordinate multi-step planning~\cite{dagan2025plancraft}, invoke tools effectively~\cite{shen2024small,lee2026don}, and produce structured outputs~\cite{xia2024fofo}. These capabilities are essential for conducting comprehensive website investigations and generating evidence-grounded reports.

However, prior work has also shown that distilling reasoning demonstrations and task trajectories from larger models can improve smaller models' task-specific reasoning and tool-use capabilities~\cite{hsieh2023distilling,chen2023fireact}. Motivated by these findings, we utilize a larger Qwen3.5-27B as a teacher or \emph{Mentor} model, and equip it with dedicated investigative prompts and tools to generate detailed demonstrations of SE scam websites' investigation and evidence-grounded reporting. We then fine-tune the Investigator and Decider models using Mentor's investigative trajectories and structured reports, aiming to transfer these specialized capabilities while preserving the computational efficiency of the smaller models. Figure~\ref{fig:framework} illustrates the framework of our study. Our study proceeds as follows:
\\
\textbf{Section~\ref{data_collection}} describes the collection of our ground-truth dataset, consisting of malicious websites from seven distinct SE scam categories along with benign websites. We similarly build our test dataset that adds an additional five categories. The former is used to generate investigative reports using our Mentor model, which are then used to train the Investigator and Decider models in our ReDence framework. The test set, on the other hand, is used to evaluate the detection and investigative categories of the ReDence framework, and also identify whether it generalizes to scam categories not covered in our ground-truth. 

This is followed by our Agentic setup in \textbf{Section~\ref{agentic-setup}}, which introduces the dedicated investigative prompts and tools we developed that allow the Mentor model, and subsequently ReDence, to have 
domain-specific knowledge on how to investigate SE scams as well as creating evidence-grounded reports. Then, we utilize the Mentor model to investigate the ground-truth websites and generate investigation trajectories and reports. We measure its detection effectiveness and manually evaluate the quality of generated reports. A case study also demonstrates how the Mentor applies the prompts and tools to gather evidence, verify claims, and substantiate its assessment. We also discuss the limitations encountered during these investigations that further motivate the need for our hybrid architecture in ReDence. 

In Section \textbf{\ref{sec:ReDence}}, we introduce the ReDence architecture, presenting the division of responsibilities between the Investigator and Decider models, and how they communicate. We then transform Mentor's trajectories and reports into complementary training datasets and finetune both these models, with Investigator learninng investigative reasoning, reporting, and delegation, while Decider learns to resolve contextual binary and multiple-choice questions. 

Finally, \textbf{Section \ref{sec:evaluation}} compares ReDence with Mentor and existing scam detectors, examining detection effectiveness, investigation time, memory requirements, and investigation and report quality. A comparative case study further illustrates how differences in exploration, interaction, and external verification affect the evidence collected and the resulting assessments.



\section{Building the SE scam corpus}
\label{data_collection}

Evaluating our agentic framework requires direct interaction with websites, including manipulating page elements and following internal links. Existing SE scam datasets primarily provide URLs \cite{mamun2016detecting}, static screenshots \cite{abdelnabi2020visualphishnet}, or pre-extracted website features \cite{prasad2024phiusiil}, which do not preserve the interactive environment required for our evaluation.
\newline
\textbf{Ground-truth websites:} We therefore construct our own ground-truth dataset by collecting candidate websites reported to the Better Business Bureau (BBB) Scam Tracker~\cite{BBB} between June 1 and July 6, 2026. BBB Scam Tracker is a repository of reported scams that has been widely used in prior research on online fraud~\cite{ma2025psyscam,maher2025fraud,gomez2025clean}. We collect 2,175 reports that contained 728 unique domain links belonging to seven most popular scam categories on BBB: online purchase scams, counterfeit product scams, advance-fee loan scams, investment scams, debt collection scams, employment scams, and sweepstakes/lottery scams. The distribution of these samples is provided in Table~\ref{tab:se_categories}, and the descriptions of these categores are taken from the BBB Scam Glossary~\footnote{https://scamsurvivaltoolkit.bbbmarketplacetrust.org/scam-type-glossary/}. The websites are validated by two independent coders: Coder 1 has research expertise in online fraud and abuse, while Coder 2 has prior experience in computer security through research and academic coursework. 

We also collect and manually validate 1,200 benign websites from Curlie \footnote{https://curlie.org/}, a web directory that organizes websites by topic, to identify legitimate services and workflows corresponding to the SE scam categories. These include online retail and related services (379), branded-product storefronts (127), lending and credit services (242), investment and wealth-management services (103), employment and recruitment services (199), and contests and prize promotions (150).
\newline
\textbf{Test set:} We collect and manually validate 2,726 reports from BBB between July 12 and August 14, 2026. These reports provide an additional 1,551 websites, out of which 1,243 belong to the seven SE categories, while the remaining 308 websites belong to five held-out categories not included in our ground truth: phishing, travel/vacation/timeshare, government agency impersonation, home improvement, and fake invoices/supplier bills. 
We use this dataset to evaluate ReDence, as well as existing ML-based scam detectors, in Section \ref{sec:evaluation}. The inclusion of held-out categories also allows us to assess how well ReDence generalizes to scam types not encountered during its training. Table~\ref{tab:se_categories} provides the distribution of this test set.

Similar to the ground truth, we also collect and manually validate an additional 1,500 benign websites from the Curlie directory, covering online retail and related services (900), branded-product storefronts (150), lending and credit services (125), investment and wealth-management services (100), employment and recruitment services (175), and contests and prize promotions (50).


We specifically construct a separate test set as our ground-truth dataset is relatively small and exhibits substantial class imbalance across scam categories, reflecting the uneven prevalence of different scam types in the real world. Thus, a train: test split for training ReDence would not have left enough samples to evaluate it. Having a seperate test set allows ReDence to learn from the full ground-truth report, and subsequently evaluate it on a new independent set of websites.
\newline
\textbf{Website preservation:} While ideally our framework would be able to investigate live scam websites, malicious websites can be ephemeral \cite{lee20257}, taken down by the hosting providers \cite{moore2007examining}, or deactivated by the attackers themselves \cite{barron2019now}.
Thus, we assume a possibility of some of the websites in our ground truth no longer being active when we run our framework. Therefore, for each website, we create highly functional, replayable snapshots using \emph{Browsertrix}\cite{browsertrix}, an open-source web-archiving system that has been widely used for high-fidelity web preservation ~\cite{sucho, untwebarch}. Browsertrix captures modern websites and packages their associated resources into WACZ archives \cite{webrecorderwacz}. These snapshots serve two purposes: i) to provide a fallback environment that the agent can look through if the original website is no longer accessible, and ii) to preserve the websites such that our results can be reproducible in the future. For each website, we configure Browsertrix to capture the entry page and pages reachable within two link traversals on the same hostname. This includes pages linked directly from the entry page and those reached by following one additional link. We cap each crawl at 100 pages. However, we find that 80.32\% of the websites in our ground truth have no more than 93 linked pages, with the median number of linked pages being 13.
To evaluate the quality of our snapshots, we utilize the \emph{Memento Damage} metric proposed by Brunelle et al. \cite{brunelle2015not}, an established measure of web-archive quality that has been adopted in subsequent studies. The metric assigns each archived webpage a normalized damage rating, $D_m \in [0,1]$, based on the relative importance of resources that are missing during replay, with lower values indicating higher-fidelity captures. Across our archived websites, we observe a median damage rating of 0.08 (IQR 0.03-0.15). 
Across our evaluations, only 373 (\~14\%) websites in our ground truth and test set became unavailable, requiring ReDence to investigate their preserved snapshots. Among these, credential-phishing websites accounted for the majority of unavailable sites (224). Although website unavailability does not always imply enforcement action, this provides an early indication that phishing campaigns might be more short-lived than other scam types. 

\begin{table}[t]
    \centering
    \small
    \setlength{\tabcolsep}{6pt}
    \renewcommand{\arraystretch}{1.12}
    \resizebox{0.8\columnwidth}{!}{%
    \begin{tabular}{@{}lrr@{}}
        \toprule
        \textbf{SE Scam Category} &
        \textbf{Ground Truth} &
        \textbf{Test Set} \\
        \midrule
        \multicolumn{3}{@{}l}{\textit{Ground-truth Categories}} \\
        Online Purchase & 587 & 637 \\
        Counterfeit Products & 109 & 203 \\
        Advance-Fee Loans & 30 & 51 \\
        Investment & 26 & 59 \\
        Debt Collection & 37 & 53 \\
        Employment & 32 & 201 \\
        Sweepstakes/Lottery/Prizes & 15 & 39 \\
        \midrule
        \multicolumn{3}{@{}l}{\textit{Held-Out Categories}} \\
        Phishing & -- & 113 \\
        Travel/Vacation/Timeshare & -- & 61 \\
        Government Agency Impersonation & -- & 34 \\
        Home Improvement & -- & 64 \\
        Fake Invoices/Supplier Bills & -- & 36 \\
        \midrule
        \textbf{Total unique websites} & \textbf{728} & \textbf{1,551} \\
        \bottomrule
    \end{tabular}%
    }
    \caption{Distribution of websites across the ground-truth
    and test datasets. Held-out categories are included only in the test set.}
    \label{tab:se_categories}
\end{table}

\section{The Mentor Model}
\label{agentic-setup}

We select Qwen3.5:27B as our \emph{Mentor} model, as it offers multimodal capabilities that allow it to interpret textual and visual observations, while its tool-calling interface enables it to invoke externally implemented tools. Moreover it offers a strong balance between instruction-following capability, computational efficiency, and memory footprint, while matching or outperforming substantially larger variants on several relevant benchmarks, including IFEval~\cite{zhou2023instruction}, IFBench~\cite{pyatkin2026generalizing}, HLE with Tools~\cite{Buniatyan2025humanitysexamwithtools}, and WideSearch~\cite{wong2026widesearch}. Its native context window of approximately 262K tokens accommodates the webpage content, tool outputs, external verification results, and evidence accumulated during multi-step investigations. 

\begin{figure}[t]
    \centering
    \includegraphics[width=\columnwidth]{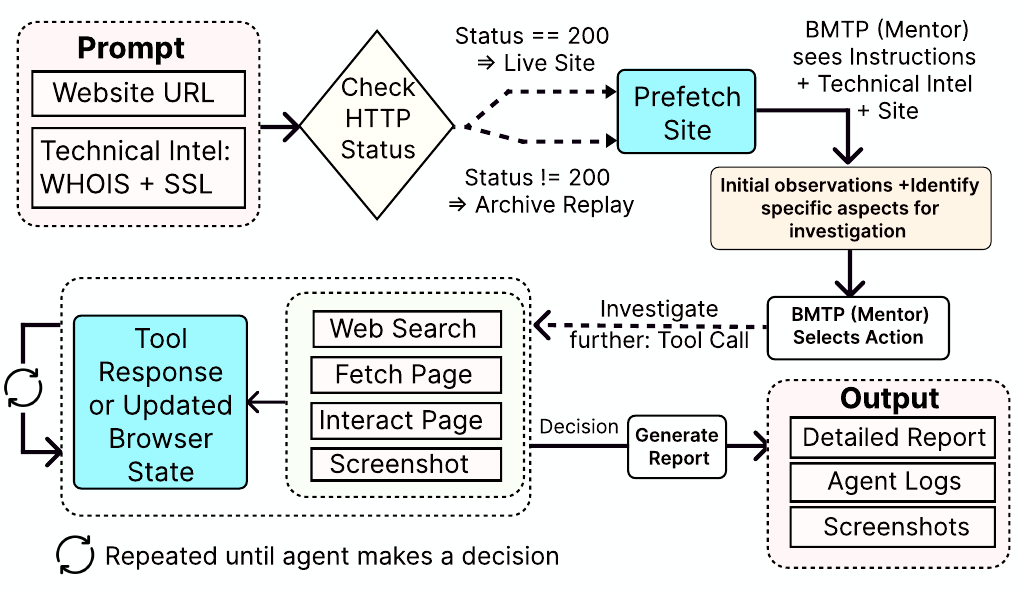}
    \caption{Overview of Mentor's investigative workflow.}
    \label{fig:agent_workflow}
\end{figure}

However, these capabilities do not inherently provide access to the internet or the ability to investigate live websites. When provided with a website URL, the base model acknowledges this limitation and offers general advice on how to identify malicious URLs. Thus, to enable such investigations, we augment the model with tools for retrieving webpage content, interacting with websites, and searching external sources, along with task-specific guidance prompts for conducting investigations and documenting evidence. Figure~\ref{fig:agent_workflow} illustrates Mentor's investigative workflow.



\subsection{Investigative Prompt}
\label{sec:promptdev}



Investigating an SE scam requires more than access to website content and external information. Analysts must decide which claims warrant verification, determine what additional evidence is needed, reconsider their assessment as new findings emerge, and document the basis for their conclusions. Enabling an agent to perform these tasks requires guidance on both the investigative process and how its findings should be reported. Prior work has found general-purpose models often lack the domain-specific knowledge or procedural expertise needed to carry out these tasks systematically~\cite{yang2023empower,fei2024lawbench,ye2025sop,jiang2026generalist}. We therefore develop a task-specific investigative prompt that guides the agent through these decisions and specifies how to document the resulting investigation. This approach builds on prior work showing that explicit instructions and procedural guidance can support specialized tasks without modifying the model's underlying parameters~\cite{nori2023can,xu2023expertprompting,ye2025sop,chen2025prompt}.
\\
We organize the prompt around four questions that connect website exploration with an evidence-grounded assessment that an analyst can independently examine:
\begin{enumerate}
    \item What does the website present, and what additional information becomes available through navigation and interaction?
    \item Which claims require verification, and what investigative actions could resolve the outstanding questions?
    \item How should newly collected evidence change the agent's assessment and subsequent actions?
    \item How should the findings, interpretations, and supporting sources be documented to support analyst review?
\end{enumerate}

These questions motivate four corresponding components: systematic website exploration, purpose-driven planning, adaptive investigation, and evidence-grounded reporting. This design draws on research on task decomposition, prompt optimization, and reasoning guidance~\cite{habba2025promptsuite,wang2024promptagent,liu2026comprehensive,wei2022chain,kojima2022large}. The prompt components are described below. The agent carries out the resulting investigative actions using the tools described in Section~\ref{investigation_tools}.


\textbf{Preamble:} We provide the agent with definitions of all 31 scam categories listed in the BBB Scam Type Glossary \cite{bbb-scam-recovery-toolkit-2025} and instruct it to determine whether a website is malicious and, if so, assign it to one of these categories. We include category descriptions to establish a consistent classification scheme and clarify distinctions between scam types, following prior work on explicit category definitions for zero-shot and fine-grained classification~\cite{peskine2023definitions,mu2305navigating}. 

\noindent\textbf{Systematic website exploration (EXP):}
The first component guides the agent in gathering observations about the website's content and behavior. A landing page may reveal only part of a website's activities; navigation and interaction can expose additional evidence, including credential-collection workflows, evasive redirects, and content behind interaction barriers~\cite{liu2022inferring,liu2023knowledge,teoh2024phishdecloaker,bitaab2023beyond}. The prompt therefore directs the agent to inspect relevant website components, follow internal links, and interact with page elements as needed before reaching a conclusion. Table~\ref{tab:p2_components} summarizes the components covered by this guidance. Initial exploration establishes the website's apparent activities and claims, providing the observations needed to formulate a more targeted investigation.

\begin{table}[t]
\centering
\caption{Website components explicitly considered in our \emph{EXP} prompt component.}
\label{tab:p2_components}
\begin{tabularx}{\columnwidth}{lX}
\toprule
\textbf{Category} & \textbf{Components} \\
\midrule
Navigation 
& Header, footer, menus, links, and redirects \\
Identity 
& Branding, logos, and claimed organization \\
Content 
& Main text, offers, representations, products, and services \\
User interaction 
& Forms, requested information, buttons, registration, login, and account access \\
Organization details 
& Contact information, policies, legal pages, and company information \\
Transactions 
& Payment methods and payment-related claims \\
\bottomrule
\end{tabularx}
\end{table}

\noindent\textbf{Purpose-driven investigative planning (PLAN):}
Building on these initial observations, the agent identifies the website's apparent purpose, claimed operator, actions requested from users, and plausible explanations for its behavior. It then identifies unresolved questions and constructs a concise plan linking each question to an investigative action that could resolve it. This follows security-investigation practices in which analysts formulate hypotheses or questions and identify the evidence needed to assess them~\cite{wafula2019carve,araujo2021evidential}. 

\noindent\textbf{Adaptive investigation (ADAPT):}
The initial plan must remain responsive to what the agent discovers. After each action, the prompt requires the agent to assess the resulting observation, reconsider its interpretation of the website, and determine whether the remaining plan should change. This follows the principle that investigative findings should inform both the plausibility of existing hypotheses and the selection of subsequent actions~\cite{tsigkanos2014engineering,pasquale2016adaptive}.

Exploration and planning continue in a loop: an unresolved question motivates an action, the resulting observation informs the assessment, and remaining or newly identified uncertainties guide the next action. The prompt prioritizes actions that resolve meaningful uncertainties and discourages exhaustive interaction without a clear investigative purpose.

\noindent\textbf{Evidence-grounded reporting (REPORT):}
The final component specifies how the agent documents its investigation and substantiates its assessment. Throughout the investigation, the agent records the initial plan and, for each subsequent step, the goal, action, observation, evidence update, and plan update. This record connects the investigative objectives with the actions performed and the findings that influenced subsequent decisions, following prior work on making language-agent behavior interpretable through explicit reasoning and action trajectories~\cite{yao2022react}.

The agent also maintains an evidence ledger that assigns each important finding an \emph{Evidence ID} and records its source, evidence type, status as a direct observation or inference, and relevance to the assessment. The ledger includes both supporting and contradictory findings. Every material factual claim in the final report must reference one or more Evidence IDs, allowing an analyst to trace the claim to its underlying evidence and examine whether the interpretation is justified. These requirements address the risk that LLM-generated reports contain assertions that are unsupported by, or contradict, the available information~\cite{niu2024ragtruth,min2023factscore}.

The final output follows a structured JSON format containing the URL, category, confidence, and explanation, together with the investigative record and evidence ledger. This provides a consistent representation for parsing, comparing, and inspecting reports across websites and runs~\cite{xia2024fofo,chen2025prompt}. The confidence field records the agent's expressed certainty; we do not treat it as a calibrated probability, since LLMs can produce highly confident but incorrect assessments~\cite{mielke2022reducing,groot2024overconfidence}.

\noindent\textbf{Additional information:}
To support the interpretation of time-dependent evidence such as (but not limited to) domain age, claims about an organization's operating history, or validity of promotions, we provide the current date. Doing so also reduces the risk of the model's confusing its knowledge cutoff with the present.

\subsection{Investigation Tools}
\label{investigation_tools}

To allow the agent to dynamically execute its investigative plan as directed by the prompt, it needs to directly interact with the website and its supporting artifacts to collect evidence and pursue further checks as new questions emerge. We implement three complementary tools to enable this: \emph{Fetch Page}, \emph{Interact Page}, and \emph{Web Search}, which provide the functionality needed to carry out the procedures specified by the investigative prompt.

\noindent\textbf{Fetch Page:}
Assessing a website requires access to both its underlying implementation and the content presented to visitors. Fetch Page retrieves four complementary representations of a webpage: its HTML source, textual content, Document Object Model (DOM), and screenshots of its rendered appearance.

The HTML source and DOM expose the page's structure, elements, and attributes, including links, forms, and embedded content. The text allows the agent to examine service descriptions, organizational claims, contact details, and payment or refund policies. Screenshots provide visual context, including branding, layout, and information embedded in images. These representations allow the agent to examine both what the website communicates and how its content is implemented.


\noindent\textbf{Interact Page:}
Some evidence becomes available only after a visitor opens a menu, selects a service, fills a form, or progresses through a registration or checkout workflow. Interact Page enables the agent to examine this behavior through a stateful browser environment implemented using Playwright~\cite{microsoftplaywright}.

We expose interactive elements through Set-of-Marks-style identifiers~\cite{yang2023set}. The agent receives a numbered representation of the available elements and uses their identifiers to select targets for actions such as clicking a button or filling a field. It can also issue keyboard actions. This avoids requiring the model to generate CSS selectors or XPath expressions, which can be brittle across websites~\cite{firecrawl}. The element identifiers are refreshed as the browser state changes, allowing subsequent actions to target the updated interface. We also integrate a CAPTCHA solver based on a Selenium library~\cite{seleniumcaptchasolver} to process supported challenges that interrupt website interaction.

Fetch Page and Interact Page operate in an observation--interaction loop. Fetch Page retrieves the current page data, the agent examines these observations and selects an action, and Interact Page executes that action on the chosen elements. The agent can assess the outcome and determine what to do next. Together, they support investigation of multi-step workflows and connect the website's initial claims with the behavior encountered as a visitor proceeds through its services.

\noindent\textbf{Web Search:}
A website's own content may provide claims without sufficient evidence to verify them. Establishing an organizational identity, professional affiliation, or business registration often require information from independent sources. Web Search supports this verification by submitting model-generated queries to web-search engines through DDGS~\cite{deedy}, an open-source metasearch library, and returning relevant search results.

The agent formulates queries using claims and artifacts encountered during its investigation, including company names, registration numbers, email addresses, telephone numbers, and distinctive phrases. These queries can help locate authoritative records, identify contact details or content reused across websites, and uncover independently reported experiences associated with the claimed operator. Because the queries follow the findings collected during the investigation, the agent can pursue the specific uncertainties raised by each website.

Web Search works with Fetch Page to move from source discovery to source inspection. The agent selects relevant result URLs and passes them to Fetch Page to examine their contents, allowing it to compare the website's representations with corroborating or contradictory external evidence.

\subsection{Evaluation}
\label{evaluation-mentor}

Utilizing this setup, the Mentor model is run on 728 malicious and 1,200 benign websites obtained from the Ground-truth dataset. In the following paragraphs, we identify its detection effectiveness (as well as issues that cause misdetection). This is followed by a case study that demonstrates how the model investigated an SE scam website and generated it's report. Finally, we follow this by emperically identifying the quality of its generated evidence reports.

\subsubsection{Detection}
The Mentor model identified 589 websites out of the 728 as malicious (80.9\%), spread across the seven SE scam categories: online purchase (482), counterfeit products (95), advance-fee loans (25), investment (22), debt collection (21), employment (27), and sweepstakes/lottery/prizes (9). It misdetected 42 websites as being benign, and 97 as having "insufficient" evidence.   

To further identify the misdetections, we systematically examine Mentor’s investigation logs for each website, including its recorded reasoning steps, tool selections, tool execution times, and how it adapted its investigation strategy and made decisions as evidence accumulated. Not constrained by a fixed time limit, Mentor required a median of 199 seconds and 17 tool calls per website. \emph{However}, we find that 68\% of the aggregate runtime was spent on decision-making tasks, including selecting tools, choosing which pages or search results to inspect, assessing how new evidence affects the investigation, and deciding whether to continue, revise, or conclude the investigation.

We also find that in several cases, Mentor repeatedly reconsidered unresolved questions and issued additional tool calls without obtaining evidence that resolved them. Among the 97 websites marked as having inconclusive evidence, we identify this phenomenon in 72 cases. This pattern also occurred in 26 of the 42 websites classified as benign, where Mentor instead treated the supposed lack of evidence as grounds for a benign verdict. Prior research shows that extended reasoning can produce redundant solutions with limited additional benefit~\cite{chen2025not}, while agent loops can repeat actions that return previously available observations without advancing the task~\cite{chen2026deloop}. We also examine exact duplicate tool calls, excluding fetches of the original target URL. These repetitions occur in 195 investigations (26.8\%). 

Together, these findings motivate integrating a decision model into ReDence to reduce decision-making overhead and limit repeated deliberation and tool calls that do not advance the investigation.

\subsubsection{Case Study} To illustrate how Mentor investigates a website’s claims and documents the evidence supporting its assessment, we examine \textsf{publictradeinc.com}, an investment scam website.
\newline
\textbf{Investigation:} Mentor begins by examining the landing page of the website using the \emph{Fetch Page} tool and finds that it advertises trading services and encourages visitors to create an account on the platform. Following the prompt's guidance on \emph{systematic website exploration}, the agent uses \emph{Interact Page} and subsequent \emph{Fetch Page} calls to inspect the website's organizational information, contact details, FAQ, and registration page.

Attempts to retrieve the legal-and-regulation, about-us, terms-and-conditions, and privacy-policy pages return 404 errors. These failures leave several questions about the operator and its terms unanswered. The contact page, however, lists offices in New York, London, and Tokyo, together with a telephone number. The agent uses \emph{Web Search} to check the advertised address and telephone number but does not find results establishing their association with PublicTrade.

These unresolved identity claims direct the investigation toward more specific information about the business. The FAQ provides a concrete lead: the website claims to be registered in the United Kingdom under company number 08683932. For a platform encouraging users to invest, checking who operates the service is central to establishing its credibility~\cite{fca2026firmchecker}. This focus reflects the prompt's \emph{purpose-driven planning}, with the website's advertised activity determining which claims require verification.

Examining the registration page reveals fields requesting a username, email address, password, and wallet address but no identity documents or other Know Your Customer (KYC) verification. Mentor flags this omission as suspicious, stating that ``legitimate brokers require identity verification,'' and contrasting this requirement with the platform's minimal registration fields. This concern aligns with UK customer due diligence requirements, which require covered financial institutions to identify customers and verify their identities~\cite{uk2017moneylaundering}. 

Mentor then searches for the company registration number. The results reveal that the same number appears on two other trading websites: \textsf{swiftapexcapital.com} and \textsf{bitcloven.com}. This leads to the model raising a further question: which company does the number identify, and is PublicTrade associated with it?

Because the website claims UK registration, the agent checks the domain's WHOIS record, identifying that it had been registered during December 2025 in Nairobi, Kenya. Subsequently, it searches for the corresponding registration number at Companies House \cite{companieshouse}, the UK government agency responsible for maintaining the public register of companies. This leads Mentor to a record showing that the number belongs to ALLBIT LTD, whose listed business activity is retail sales through mail order or the internet, \emph{a mismatch with the cryptocurrency trading and investment services advertised by \textsf{publictradeinc.com}.} Thus, neither the company name nor its listed activity substantiates PublicTrade's advertised financial-service identity. A further search combining PublicTrade and ALLBIT LTD also fails to establish a relationship between them. Mentor interprets these findings as indicating that the platform is using another company's credentials to establish credibility.

To further validate its suspicions, Mentor searches for the domains associated with the same company registration number, namely \textsf{swiftapexcapital.com} and \textsf{bitcloven.com}, and identifies that they had been previously flagged by scam-tracking websites \textsf{scamtracinghub.com} and \textsf{trackfrauds.com}, and have multiple Trustpilot reviews of users being scammed.

This overall progression illustrates Mentor's \emph{adaptive investigation} capabilities: the registration number claim prompts an external search, its reuse introduces another uncertainty, and the agent follows up by checking the official record and the claimed association. The investigation concludes by bringing together the unresolved corporate identity, reused registration number, trading-bot profit claim, unavailable legal pages, advertised payment methods, and absence of identity verification during initial registration. Mentor interprets this combination as a platform using unsubstantiated legitimacy claims to encourage investment and classifies the website as an \textbf{investment or cryptocurrency scam}, with \textbf{high confidence}.

\begin{figure}[t]
    \centering
    \includegraphics[width=\columnwidth]{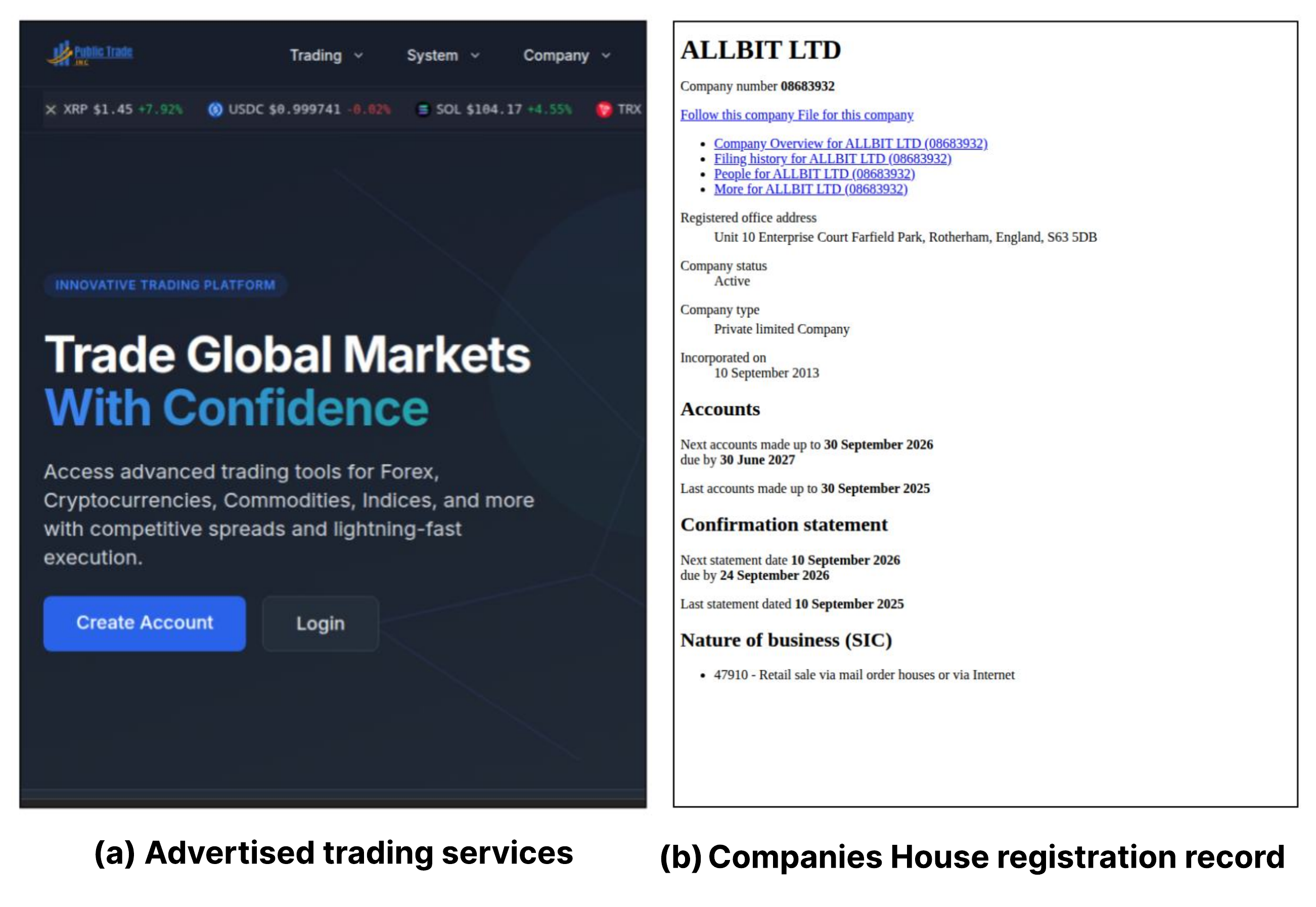}
    \caption{PublicTrade’s website (a) The landing page advertises cryptocurrency, forex, and other trading services. (b) The Companies House record for the website’s claimed registration number lists retail sales as its business activity, a mismatch with its claimed purpose.}
    \label{fig:mentor_case_study}
\end{figure}

\begin{figure}[t]
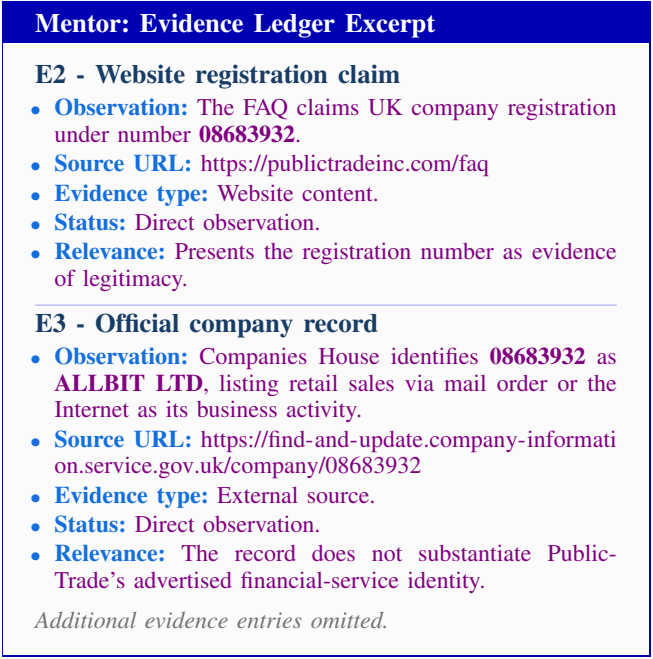

\centering

\begin{tcolorbox}[
    enhanced,
    colback=gray!3,
    colframe=blue!70!black,
    boxrule=0.5pt,
    fonttitle=\bfseries,
    title={Mentor: Evidence Ledger Excerpt},
    colbacktitle=blue!70!black,
    coltitle=white,
    toptitle=4pt,
    bottomtitle=2pt,
    sharp corners,
    boxsep=0pt,
    left=12pt,
    right=12pt,
    top=8pt,
    bottom=10pt,
    width=\columnwidth,
    unbreakable
]

{\small

\textcolor{ledgernavy}{\normalsize\textbf{E2 - Website registration claim}}

\begin{itemize}[
    leftmargin=0.8em,
    label=\textcolor{numberblue}{\scriptsize$\bullet$},
    itemsep=1pt,
    topsep=2pt,
    parsep=0pt
]
    \item \textcolor{numberblue}{\textbf{Observation:}}
    \textcolor{purpletext}{The FAQ claims UK company registration under number \textbf{08683932}.}

    \item \textcolor{numberblue}{\textbf{Source URL:}}
    \textcolor{purpletext}{\nolinkurl{https://publictradeinc.com/faq}}

    \item \textcolor{numberblue}{\textbf{Evidence type:}}
    \textcolor{purpletext}{Website content.}

    \item \textcolor{numberblue}{\textbf{Status:}}
    \textcolor{purpletext}{Direct observation.}

    \item \textcolor{numberblue}{\textbf{Relevance:}}
    \textcolor{purpletext}{Presents the registration number as evidence of legitimacy.}
\end{itemize}

\smallskip
{\color{blue!35}\hrule}
\smallskip

\textcolor{ledgernavy}{\normalsize\textbf{E3 - Official company record}}

\begin{itemize}[
    leftmargin=0.8em,
    label=\textcolor{numberblue}{\scriptsize$\bullet$},
    itemsep=1pt,
    topsep=2pt,
    parsep=0pt
]
    \item \textcolor{numberblue}{\textbf{Observation:}}
    \textcolor{purpletext}{Companies House identifies \textbf{08683932} as \textbf{ALLBIT LTD}, listing retail sales via mail order or the Internet as its business activity.}

    \item \textcolor{numberblue}{\textbf{Source URL:}}
    \textcolor{purpletext}{\nolinkurl{https://find-and-update.company-information.service.gov.uk/company/08683932}}

    \item \textcolor{numberblue}{\textbf{Evidence type:}}
    \textcolor{purpletext}{External source.}

    \item \textcolor{numberblue}{\textbf{Status:}}
    \textcolor{purpletext}{Direct observation.}

    \item \textcolor{numberblue}{\textbf{Relevance:}}
    \textcolor{purpletext}{The record does not substantiate PublicTrade's advertised financial-service identity.}
\end{itemize}

\smallskip
{\color{black!55}\textit{Additional evidence entries omitted.}}

}

\end{tcolorbox}

\caption{Condensed excerpt from Mentor's PublicTrade evidence ledger, documenting the website's registration claim and the corresponding official company record.}
\label{fig:mentor_evidence_excerpt}
\end{figure}

\textbf{Reporting:} The generated report documents the website’s purpose, claimed organization, verdict, and confidence, followed by an evidence ledger containing twelve findings. Each entry includes an Evidence ID, observation, source URL, evidence type, status, and explanation of its relevance. The complete report references these entries in its key-evidence section and final explanation.

Figure~\ref{fig:mentor_evidence_excerpt} presents two entries illustrating how the ledger supports verification of the website’s corporate identity. E2 records the FAQ’s claim of UK company registration under number 08683932 and identifies the page containing that claim. E3 records the corresponding Companies House finding, which identifies ALLBIT LTD and lists retail sales via mail order or the Internet as its business activity. Its relevance field explains that these details do not substantiate PublicTrade’s advertised financial-service identity. Together, the entries allow an analyst to locate the website’s claim, inspect the official record, and assess whether the reported interpretation is justified.

\subsection{Evaluating Reports}
\label{evaluating_reports}

\noindent\textit{Note: We are currently repeating this analysis with two independent coders to measure intercoder agreement. The evaluation described below uses one coder.}

We now evaluate whether Mentor's reports provide relevant, verifiable evidence that analysts can use to assess a website's legitimacy. Our evaluation considers both whether the coder correctly classifies websites and whether the reports reveal additional evidence, including for websites the coder classifies correctly. We focus on evidence coverage and auditable evidence coverage because analysts need findings that substantiate an assessment and sufficient documentation to independently verify them~\cite{kramer2025integrating,rashkin2023measuring,gao2023enabling}.

Our coder is a cybersecurity professional specialized in identifying online fraud and abuse, including SE scam websites. 
We select five malicious websites from each of the seven SE scam categories, yielding 35 malicious websites. For each category, we also select five benign websites offering corresponding products or services, producing a balanced sample of 70 websites. Our coder investigated each website for 90 seconds, recording a classification and the distinct findings supporting or contradicting its legitimacy. This record is finalized before the coder sees Mentor's report. The coder subsequently examines the report to determine whether it recovers the initial findings, identifies additional relevant evidence, and provides sufficient supporting material to verify its claims.

For each website $w$, let $H_w$ denote the deduplicated set of correct, relevant findings recorded during the initial manual investigation, and let $M_w$ denote the correct, relevant findings identified in Mentor's report. The coder checks findings from both sources against the website, external sources, or collected artifacts. Additional findings from Mentor are accepted only after verification. Repeated descriptions of the same finding are counted once, and evidence supporting legitimacy is considered alongside evidence suggesting deception.

We construct a pooled reference set $G_w = H_w \cup M_w$, representing the verified findings discovered through either process. This is an observed evidence inventory rather than an exhaustive account of everything that could be discovered about the website. We calculate two metrics:

\noindent\textbf{Evidence coverage (EC):}
The proportion of verified reference findings correctly identified by Mentor:
\begin{equation}
    \mathrm{EC}(w)
    =
    \frac{|M_w|}{|G_w|}.
    \label{eq:evidence_coverage}
\end{equation}
This metric credits both findings initially identified by the coder and additional findings uncovered by Mentor. Evaluating distinct findings follows prior work on reference-based content coverage~\cite{nenkova2004evaluating}.

\noindent\textbf{Auditable evidence coverage (AEC):}
The proportion of verified reference findings that Mentor both identifies and documents sufficiently for independent verification:
\begin{equation}
    \mathrm{AEC}(w)
    =
    \frac{\sum_{e \in M_w} a_w(e)}{|G_w|},
    \label{eq:auditable_evidence_coverage}
\end{equation}
where $a_w(e)=1$ if the report provides an identifiable source, collected artifact, or documented interaction sufficient to verify finding $e$, and $0$ otherwise. A source URL alone is insufficient if its contents do not substantiate the finding. This criterion reflects the importance of source-supported claims in attributable reporting~\cite{rashkin2023measuring,gao2023enabling}.

The coder correctly identifies 21 of the 35 malicious websites, incorrectly classifies 7 as benign, and returns inconclusive assessments for the remaining 7. Among the 35 benign websites, the coder correctly classifies 31 and incorrectly labels 4 as malicious. These outcomes illustrate the difficulty of distinguishing legitimate services from deceptive websites within a brief manual investigation. Correct classification also does not necessarily imply a comprehensive investigation. Among the 52 websites the coder classifies correctly, the initial manual records contain an average of 2.4 distinct relevant findings per website, compared with 6.8 in Mentor’s reports for the same subset. Thus, even for correctly classified websites, Mentor's reports provide additional evidence supporting the assessment.

Across the full sample, Mentor achieves a mean evidence coverage of 0.87 and auditable evidence coverage of 0.79. Its reports contain 364 additional findings absent from the coder’s initial records, of which 327 are treated as auditable.

Overall, these findings demonstrate that Mentor's reports not only go beyond providing a classification decision by uncovering relevant evidence that analysts may overlook during time-constrained manual investigations, but also enrich an analyst's understanding of the attack, even when the analyst's initial classification is correct. 
\section{ReDence}
\label{sec:ReDence}

To develop \emph{ReDence}, we distill the Mentor model's investigative and reporting capabilities into two smaller models that work together with complementary responsibilities: 1) \emph{Investigator}, based on Qwen3.5-4B, handles open-ended tasks such as planning investigations, interpreting findings, and generating evidence-grounded reports; and
2) \emph{Decider}, based on Tev1-4B, handles binary and multiple-choice decisions, including selecting browser actions, assessing evidence against specified criteria, and choosing subsequent investigative steps. Decision models such as Tev1 are designed to rapidly select among candidate options given structured context and a question~\cite{holmes2026decisionmodels}. This makes them particularly well-suited for frequent, well-defined decisions, where traditional LLMs incur substantial computational and latency overhead from autoregressive token generation~\cite{madaan2025rethinking}. 

The investigator formulates decision requests, which Decider resolves through its concise selection interface. Table~\ref{tab:investigator_Decider_roles} illustrates how responsibilities are divided between the two models across the different stages of a website investigation.

We also include \emph{Controller}, an in-between module which coordinates their exchanges, passes the relevant context and options to Decider, returns its selection to Investigator, and executes the associated tool operations. We illustrate the overall configuration, as well as how these models were fine-tuned in the proceeding sections. 



\begin{table*}[t]
    \centering
    \small
    \renewcommand{\arraystretch}{1.25}
    \setlength{\tabcolsep}{8pt}

    \resizebox{0.75\textwidth}{!}{%
    \begin{tabularx}{\textwidth}{@{}
        >{\raggedright\arraybackslash}p{0.15\textwidth}
        >{\raggedright\arraybackslash}X
        >{\raggedright\arraybackslash}X
    @{}}
        \toprule
        \textbf{Task} &
        \textbf{Investigator}\enspace{\footnotesize Qwen3.5-4B} &
        \textbf{Decider}\enspace{\footnotesize Tev1-4B} \\

        & \emph{Open-ended reasoning and generation}
        & \emph{Binary and multiple-choice decisions} \\
        \midrule

        \textbf{Planning} &
        Identify the website's purpose, define objectives,
        and formulate an investigative plan. &
        Choose among candidate actions addressing
        the current objective. \\
        \addlinespace[4pt]

        \textbf{External verification} &
        Identify claims requiring verification
        and generate search queries. &
        Select search results or external sources
        for retrieval. \\
        \addlinespace[4pt]

        \textbf{Browser interaction} &
        Define the interaction's objective
        and supply required input values. &
        Select links, buttons, form options,
        or other available browser actions. \\
        \addlinespace[4pt]

        \textbf{Evidence assessment} &
        Organize observations, interpret relationships,
        and formulate evidence questions. &
        Resolve explicit binary or multiple-choice
        questions about the available evidence. \\
        \addlinespace[4pt]

        \textbf{Adaptive follow-up} &
        Identify unresolved questions
        and formulate revisions to the plan. &
        Choose whether to continue, revise,
        or conclude the investigation. \\
        \addlinespace[4pt]

        \textbf{Final assessment} &
        Synthesize findings and generate
        an evidence-grounded report. &
        Select the verdict and applicable scam category
        from the supplied alternatives. \\
        \bottomrule
    \end{tabularx}%
    }

    \caption{Division of responsibilities between
    ReDence's Investigator and Decider.}
    \label{tab:investigator_Decider_roles}
\end{table*}

\subsection{Constructing the Training Datasets}
\label{sec:ReDence_data}

Prior work has shown that explanations and reasoning traces generated by a larger model can improve the task-specific capabilities of smaller models~\cite{hsieh2023distilling,li2025naturalthoughts,guo2025deepseek,ahmad2025opencodereasoning}. 
We construct both datasets from Mentor's investigations of malicious and benign websites in our ground-truth dataset (Section ~\ref{evaluation-mentor}). Each trajectory begins with the system prompt defining the agent's role and available tools, followed by the initial user message containing the target URL, investigation requirements, and prefetched information. The remaining conversation records Mentor's intermediate responses, tool calls, and tool observations, concluding with its structured report.

These trajectories preserve the information available when each investigative action was selected. They therefore capture the sequential nature of website investigation, where observations inform subsequent actions and assessments, following prior work that uses action--observation sequences to gather external information and guide decisions across multiple steps~\cite{yao2022react, nakano2021webgpt, chen2023fireact}.

\noindent\textbf{Investigator training data:}
Investigator receives most of the original trajectory, including Mentor's planning, open-ended interpretation, evidence synthesis, and final reporting. We modify the instructions and corresponding assistant turns to demonstrate when and how to delegate decisions to Decider through the controller. These delegated decisions include choosing a search result, selecting a website element or form option, answering a bounded question about collected evidence, and choosing whether to continue, revise, or conclude an investigation.

For each eligible decision, we replace Mentor's original selection step with a delegation request containing the investigative objective, relevant context, decision question, and available alternatives. Investigator's next training target then demonstrates how to use this returned information to continue the investigation.

\noindent\textbf{Decider training data:}
Decider is trained on individual decision examples extracted from the same Mentor trajectories. Each example contains the investigative context, a question, and a set of labeled alternatives. The target is the option corresponding to Mentor's recorded action or explicit judgment. Unlike Investigator, Decider is trained to return the selected option without generating the surrounding investigative explanation.

The selection examples cover five types of decisions:
\begin{itemize}
    \item Choosing which search result
    or external source to retrieve for a specified verification task.

    \item Choosing among
    available links, buttons, form options, or other interactions
    relevant to the current objective.

    \item Answering bounded questions
    about whether the available observations support, contradict,
    or fail to establish a particular claim.

    \item Choosing whether to
    continue the current approach, revise it, or conclude
    the investigation.

    \item Selecting the final
    verdict and, when applicable, a scam category using
    the accumulated evidence.
\end{itemize}

For each example, we derive the candidate choices and correct response from Mentor's recorded investigation. For action-selection tasks, the correct response is the action Mentor performed, while for evidence-assessment tasks, it is Mentor's interpretation of the available evidence. Final-assessment tasks use the evidence accumulated before Mentor generates its concluding report. Browser-action choices are drawn from the recorded webpage state, whereas source-selection choices are derived from the search results returned during the investigation. We retain only examples where the context, candidate choices, and correct response can be reliably reconstructed from Mentor's trajectory. Where appropriate, we also include options such as insufficient evidence or continuing the investigation.

\textbf{Finetuning configuration:} 
We independently fine-tune both models using QLoRA with frozen 4-bit backbones and trainable LoRA adapters~\cite{dettmers2023qlora}, implemented through Hugging Face Transformers and PEFT \cite{peft}. Completion-only cross-entropy loss is applied to Investigator's target assistant responses and Decider's option labels and end-of-sequence markers; context and tool observations are excluded from the loss. Investigator uses a maximum sequence length of 16,384 tokens, LoRA rank 16, $\alpha=32$, and three training epochs at a learning rate of $1\times10^{-4}$. Decider uses 2,048 tokens, rank 8, $\alpha=16$, and one epoch at $5\times10^{-5}$. 


\subsection{Framework Coordination with Controller}
\label{sec:ReDence_controller}

\begin{figure}[t]
    \centering
    \includegraphics[width=0.8\columnwidth]{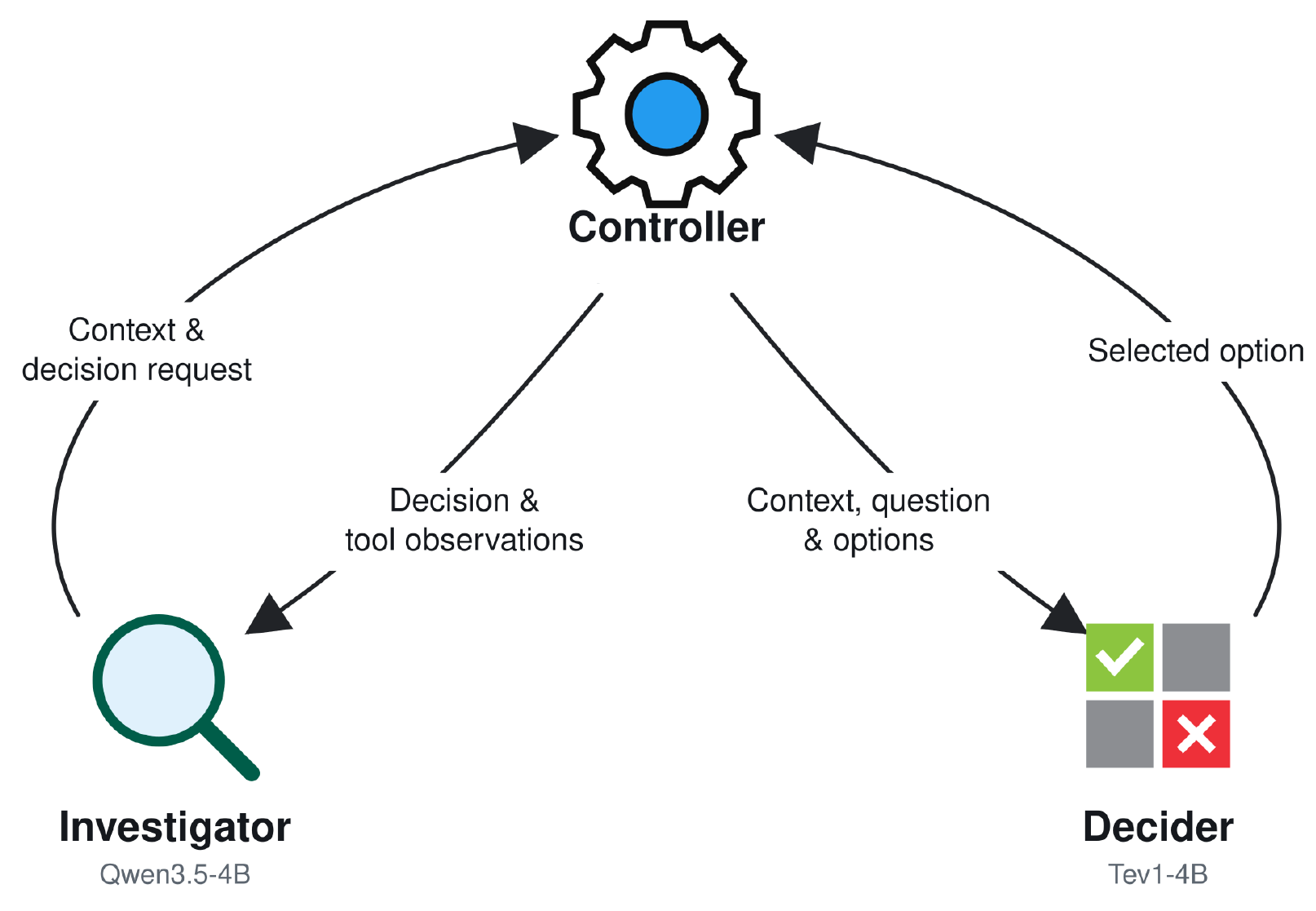}
    \caption{Controller routes context and decision requests from Investigator to Decider, and then returns decisions and tool observations back to Investigator.}
    \label{fig:controller}
\end{figure}

Controller connects the two models with the investigation tools described in Section~\ref{investigation_tools}. Investigator maintains the broader investigative context and identifies when a task requires a binary or multiple-choice decision. It then sends Controller a request containing the current objective, relevant observations and uncertainties, the decision question, and the available alternatives. Controller validates the request and passes it to Decider. Decider returns its selected option to the controller, which records the decision and passes it back to Investigator.

When the selected option specifies a tool operation, the controller executes that operation and returns its results alongside the choice. Investigator does not need to generate the same tool call again. This allows Decider to direct source retrieval or browser interaction within the task specified by Investigator, while the controller handles execution and preserves the resulting observations.

For example, Investigator may identify a navigation menu and seek the page most likely to establish the website's claimed operator. It supplies this objective and the available menu entries through the controller. Decider chooses the relevant entry, and the controller performs the interaction, retrieves the updated page content, and refreshes the available element identifiers. Investigator then receives the selected action and its outcome, interprets the new information, and continues the investigation.

The same handoff applies to evidence judgments and investigation control. Investigator formulates a specific question and supplies the observations needed to answer it; while Decider chooses among the permitted responses, and Investigator incorporates the returned judgment into its subsequent reasoning. A judgment about one feature is distinguished from an assessment of the entire website.

Controller records the context supplied for each decision, Decider's response, the tools executed, and the resulting observations or errors. Tool outputs are retained as observations, while Decider's answers are recorded as model judgments. Investigator uses these records to maintain the evidence ledger, reconsider unresolved questions, and explain how the findings relate to its assessment. Before producing the final report, Investigator synthesizes the accumulated evidence and delegates the verdict and applicable category selection through the same interface.

Figure~\ref{fig:controller} illustrates a macroscopic diagram of how each of the three components i.e. Investigator, Controller and Decider communicate with each other, whereas Algorithm~\ref{alg:investigative_prompt} summarizes this coordination across the four investigative stages. Controller also enforces the execution budget (120 seconds per website) and checks the final report's structure and evidence references.


\begin{algorithm}[t]
\caption{ReDence investigation with delegated decisions.}
\label{alg:investigative_prompt}

\begin{adjustbox}{
    max totalsize={\columnwidth}{0.6\textheight},
    center
}
\begin{minipage}{\columnwidth}
\footnotesize
\algrenewcommand{\algorithmicindent}{0.6em}

\begin{algorithmic}[1]
\Require URL, date, BBB categories, budget
\Ensure Structured assessment with traceable evidence
\State Define Investigator $I$, Decider $D$, Controller $C$;
       initialize evidence ledger $L$ and investigation record $R$

\Statex \textbf{Stage 1: Systematic website exploration}
\State $C$ fetches the initial page; if inaccessible,
       \Return access details and \emph{insufficient evidence}
\State $I$ examines text, visual content, claims,
       and available interactions

\Statex \textbf{Stage 2: Purpose-driven planning}
\State $I$ identifies purpose, operator, requested user actions,
       and plausible explanations; links unresolved questions
       to investigative objectives and candidate checks

\Statex \textbf{Stage 3: Adaptive investigation}
\While{budget remains and investigation is not concluded}
    \State $I$ prepares context, questions, candidate options,
           search queries, and interaction inputs as needed
    \State Through \textsc{Decide}, $D$ selects tools,
           pages/search results, or browser actions
           such as links, buttons, and form options
    \State $C$ executes selected or directly requested tool calls;
           returns updated observations or errors to $I$
    \State $I$ interprets findings and formulates evidence questions;
           $D$ resolves binary and multiple-choice judgments
           through \textsc{Decide}
    \State $I$ updates $L$ with evidence IDs, observations,
           sources, types, observation/inference status, and relevance
    \State Through \textsc{Decide}, $D$ selects
           \emph{continue}, \emph{revise}, or \emph{conclude};
           $I$ updates the plan and records objectives,
           decisions, actions, observations, and updates in $R$
\EndWhile

\Statex \textbf{Stage 4: Evidence-grounded reporting}
\State $I$ synthesizes supporting and contradictory findings;
       $D$ selects the verdict and applicable BBB category
       through \textsc{Decide}
\State $I$ generates JSON containing $L$, $R$, evidence-linked
       explanations, limitations, and unresolved questions
\State $C$ validates structure and evidence references;
       \Return report

\Statex
\Function{Decide}{$x,q,O$}
    \State $I \rightarrow C \rightarrow D$:
           context $x$, question $q$, options $O$
    \State $D \rightarrow C \rightarrow I$:
           selected option; $C$ validates requests and selections,
           records the outcome, and returns a decision or explicit error
\EndFunction
\end{algorithmic}

\end{minipage}
\end{adjustbox}
\end{algorithm}

\section{Evaluating ReDence}
\label{sec:evaluation}

To identify the effectiveness of the ReDence framework with respect to both speed and detection accuracy, we compare it directly with the Mentor model, as well as three existing automated detectors: URLNet~\cite{le2018urlnet}, BeyondPhish~\cite{bitaab2023beyond}, and ScamFerret~\cite{nakano2025scamferret}. 

URLNet~\cite{le2018urlnet} is a convolutional neural network that learns character-level and word-level representations directly from URL strings to identify malicious URLs. We include URLNet because URL-based features remain widely used in scam detection systems \cite{bitaab2023beyond, safi2023systematic}, primarily due to their computational efficiency and suitability for large-scale detection. However, prior research has demonstrated how scammers have increasingly adapted their URL structures and naming strategies to evade such detection approaches \cite{zhang2026anti}.

In contrast, BeyondPhish~\cite{bitaab2023beyond} uses predefined features derived from website content, URLs, domain information, and associated social media accounts to classify fraudulent websites.

Finally, ScamFerret~\cite{nakano2025scamferret} is the closest to ReDence in terms of functionality. It employs a text-based LLM guided by predefined scam indicators to iteratively select from a fixed set of tools, gather information, and explain its assessment. However, ScamFerret cannot directly interact with websites or dynamically adapt its investigation based on observed behavior, while its reliance on a fixed set of tools further constrains the evidence it can collect. As we demonstrate in Section \ref{cases-study}, these limitations cause ScamFerret to overlook important evidence and malicious workflows, reducing its detection effectiveness. Furthermore, although ScamFerret provides rationales for its classifications, it lacks an evidence ledger that explicitly links conclusions to specific observations, sources, and investigative actions while distinguishing factual observations from interpretations. In Section \ref{redence-reporting}, we further demonstrate how this limitation reduces the verifiability and practical utility of its investigation reports.

We evaluate all six configurations on the separate test set described in Section~\ref{data_collection}, comprising 1,551 scam websites and 1,500 benign websites. As both URLNet or BeyondPhish are classifier tools, they do not provide any logs or reports for their actions, and thus, we only test their detection effectiveness. On the other hand, we evaluate ScamFerret for both detection and reporting effectiveness, along with a case study (Section~\ref{cases-study}) where we directly compare the investigative approach taken by ReDence with ScamFerret. 

\begin{table*}[t]
    \centering
    \small
    \renewcommand{\arraystretch}{1.2}
    \setlength{\tabcolsep}{8pt}

    \resizebox{0.8\textwidth}{!}{%
    \begin{tabular}{@{}lcccccc@{}}
        \toprule
        \textbf{Configuration} &
        \textbf{Accuracy} &
        \textbf{Precision} &
        \textbf{Recall} &
        \textbf{F1} &
        \textbf{\shortstack{Time\\(Median, s)}} &
        \textbf{\shortstack{GPU Usage\\(Median, GB)}} \\
        \midrule

        \rowcolor{blue!10}
        ReDence (4B)
        & 0.96 (0.95) & 0.97 (0.95)
        & 0.98 (0.98) & 0.97 (0.96)
        & 43.8 (27.2) & 3.4GB \\

         Mentor (27B)
        & 0.82 (0.82) & 0.83 (0.82)
        & 0.81 (0.83) & 0.82 (0.82)
        & 194.0 (151.2) & 33.8GB \\

        \midrule

        URLNet
        & 0.41 & 0.35 & 0.19 & 0.25
        & 1.2 (1.2) & 1.5GB \\

        BeyondPhish
        & 0.72 & 0.87 & 0.70 & 0.78
        & 53.7 (1.8) & 3.7GB \\

        ScamFerret (GPT 4.1)
        & 0.79 & 0.84 & 0.81 & 0.82
        & 88.5 (41.2) & API-bound \\

        ScamFerret (Llama3.3:70B)
        & 0.71 & 0.75 & 0.67 & 0.71
        & 127.4 (92.7) & 45.7GB \\

        \bottomrule
    \end{tabular}%
    }

    \caption{Comparison of detection performance and computational
    efficiency across the models on the test set. Parenthesized
    values for detection metrics for ReDence and Mentor indicate
    performance on the \emph{held-out categories}, while those for
    median time indicate inference time \emph{excluding} website
    latency.}
    \label{tab:detection_comparison}
\end{table*}

\subsection{Detection Effectiveness}
\label{sec:detection}

Table~\ref{tab:detection_comparison} illustrates the detection effectivess of the tested tools. The values in paranthesis denote the performance on the 308 scams for the held-out categories that were not included in the ground-truth.

We find that ReDence not only matches but significantly exceeds the detection performance of the Mentor model, achieving an F1 score of 0.97 compared to 0.82 for the Mentor. A key factor contributing to ReDence's effectiveness is its ability to complete investigations for all websites in the test set, despite operating under a time constraint. On the other hand, the Mentor model failed to complete the investigation for 208 websites.

We further observe that both models retain their detection effectiveness on the held-out categories of social engineering (SE) scams, achieving F1 scores of 0.96 and 0.82 for ReDence and Mentor, respectively. Notably, ReDence maintains the same recall of 0.98 on both the development and held-out categories, with only a marginal reduction in precision. These results provide evidence that ReDence's investigative capabilities generalize beyond the scam categories represented in the ground truth categories that it was trained on.

ReDence also substantially improves computational efficiency, both in terms of runtime and memory consumption. It completes website investigations in a median of 43.8 seconds, compared to 194 seconds for the Mentor model. We also found that ReDence required significantly fewer tool calls, with a median of nine per website compared to 15 for Mentor. Decision-making also contributed to only 14\% of ReDence's aggregate runtime, compared to 71\% for Mentor, illustrating the potential reduction in overhead from delegating discrete choices to the Decider model. Similarly, ReDence consumes only 3.4 GB of GPU memory, with the CPU-bound Decider model using three additional GBs of system RAM, compared to 33.8 GB for the Mentor. 

Among the ScamFerret configurations, the GPT-4.1 implementation achieves better detection performance, with an F1 score of 0.82. The ScamFerret framework imposes a maximum of 15 iterations, where each iteration represents a cycle of LLM reasoning and tool selection or execution, or a runtime limit of 600 seconds, whichever is reached first.
Adhering to these limits, we found that it failed to complete investigations for 115 websites, during which it encountered investigation errors, prematurely terminated investigations, and incomplete reports. We also observed that evaluating this configuration on our test set alone incurred about \$39 in API charges. Considering that our test set realistically represents a small fraction of the websites that commercial security vendors may need to examine daily, this raises concerns about the cost and scalability of ScamFerret's API-based investigations.

The Llama-3.3-70B configuration of ScamFerret achieves lower detection performance, with an F1 score of 0.71. A substantial portion of this reduction is attributable to its failure to complete investigations for 239 websites,. This configuration also incurs the highest GPU memory consumption among the evaluated models (45.7 GB) and the second-longest median investigation time (127.4 seconds), behind the Mentor model. 

Among the classification-based detectors, URLNet has the lowest memory consumption (1.5 GB) and shortest runtime (1.2 seconds), but also exhibits the poorest detection performance, with an F1 score of only 0.25. This result is consistent with the limitations of relying exclusively on URL-level characteristics, as most of the malicious websites in our test set do not exhibit the suspicious URL patterns that such models are designed to recognize, being consistent with the real world where attackers presently use legitimate domain names and free SSL ceritification. Thus, even though URLNet offers substantial computational savings, it comes at a considerable cost to detection coverage.

Finally, BeyondPhish has a slightly slower median runtime (53.7 seconds), while requiring 3.7GB system RAM. Its overall detection performance is considerably lower, with an F1 score of 0.78. In particular, BeyondPhish exhibits relatively high precision (0.87) but lower recall (0.70), missing a substantial proportion of malicious websites. We also observe that BeyondPhish performs notably better on online purchase scams, identifying 522 out of 637 such websites (~82\%) in the test set. It also detected 72.4\% of scams for the counterfeit category, which is related to online purchase. However, it has significantly lower performance for other SE scam categories such as Employment scams (44.2\%) and sweepstakes (17.9\%.)

\subsection{Case Study: wego-france.com}
\label{cases-study}

We present a case study to illustrate in-depth how ReDence and ScamFerret investigate the same website, highlighting differences in the information considered, investigative actions performed, and evidence supporting their conclusions. 

\begin{figure*}[h]
    \centering
    \includegraphics[width=\textwidth]{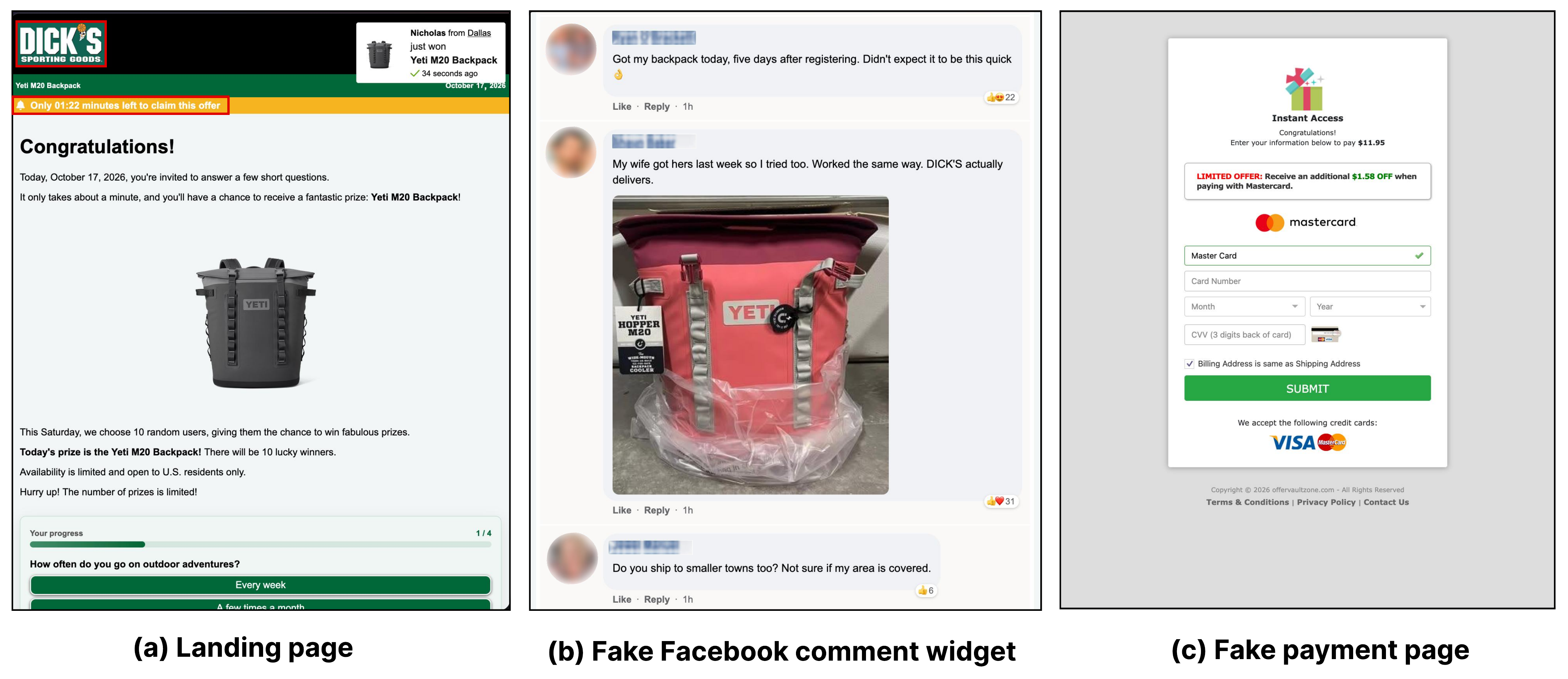}
    \caption{Preview of wego-france.com: (a) Landing page containing the organization (Dick's) logo, the (fake) timer to induce urgency, and the survey. (b) The fake facebook comment widget without any interactive elements. (c) The fake payment page which is only visible after interacting with the survey and the subsequent page which asks for non-payment information.}
    \label{fig:}
\end{figure*}

We examine \textbf{wego-france.com/backpack}, a sweepstakes scam website impersonating a DICK'S Sporting Goods giveaway. The website invites visitors to complete a survey for a chance to receive a YETI backpack, subsequently directing them to requests for personal and credit-card information. It also has a valid SSL certificate issued by Let's Encrypt, a common practice among scammers seeking to make their websites appear legitimate \cite{netcraft-2017}. By manually investigating the website, we found the following malicious artifacts:

\begin{itemize}
    \item The page displays the DICK'S Sporting Goods logo in its banner. This branding does not appear elsewhere on the website, either as an image or as text.

    \item The promotion claims that only ten winners will be selected and displays a 15-minute countdown to encourage immediate participation. However, the countdown resets to a higher value after approaching zero, while the offer remains available, undermining the claimed deadline.

    \item Facebook-style comment widgets describe successful deliveries and include photographs of supposedly received backpacks. However, the commenter names do not link to actual Facebook profiles, and the reaction buttons are nonfunctional.

    \item Completing a four-question survey reveals a gift-box game. In the observed sequence, the first selection produces an empty-box message and encourages another attempt, while the second displays a winning message and assigns a prize identifier.

    \item A payment requirement is introduced after the visitor wins. The winning screen directs visitors to a shipping page and instructs them to pay for delivery. The subsequent workflow requests their name, email address, telephone number, and delivery address, followed by credit-card details, to claim the advertised reward.
\end{itemize}

We tracked the investigative steps taken by ReDence and ScamFerret by looking at their execution logs. Both models record the selected tool, tool input, returned observation, and model’s stated interpretation before the final answer.

Both ReDence and ScamFerret begin by retrieving the website's content, but differ in their ability to identify the retailer being impersonated. ReDence uses its \emph{Fetch Page} tool to examine the website's text, DOM, and rendered appearance. By inspecting the logo image, it recognizes a visual imitation of DICK'S Sporting Goods, which does not align with the website's URL, and identifies the claimed retailer affiliation as requiring verification. ScamFerret, in contrast, retrieves the website's text, source code, and hyperlinks but fails to recognize the impersonated retailer. The logo is stored under the generic filename \textsf{assets/logo.png}, and the brand name does not appear elsewhere in the website's text or source code. Consequently, ScamFerret associates the investigated URL with Wego, identifying \textsf{wego.com} as its official domain, rather than recognizing the DICK'S Sporting Goods impersonation.

Both agents subsequently retrieve the domain's WHOIS record and identify that it was registered in May 2026, four months before the investigation. However, they interpret this information differently. ReDence acknowledges that recently registered domains are often associated with malicious activity, but recognizes that domain age alone is insufficient to establish maliciousness and redirects its attention to the website. ScamFerret, on the other hand, incorrectly rejects the WHOIS record as invalid, claiming that the registration date is in ``the future.'' We assume this results from its underlying GPT-4.1 model's June 2024 knowledge cutoff.

ReDence then examines the website's product photographs, 15-minute countdown, and Facebook-style comments, including apparent replies from a verified retailer account. Based on these observations, it formulates an investigative plan to verify the credibility of the endorsements, test the advertised deadline, and determine what visitors must do to claim the reward. ScamFerret also identifies the Facebook-style comments but cannot interact with them to verify their authenticity. While it observes that the profile names and reaction controls lack hyperlinks, it considers this evidence inconclusive, suggesting that the overwhelmingly positive reviews could reflect comments the website ``intentionally'' highlights to promote the product. ReDence, in contrast, uses \emph{Interact Page} to attempt to follow the commenter identities and activate the reaction controls. It confirms that the profile names are not clickable and the reaction buttons are nonfunctional. Combined with the overwhelmingly positive comments, these observations lead ReDence to identify the section as fabricated review engagement promoting a counterfeit product.

ReDence further investigates the 15-minute countdown associated with the advertised prize. Rather than accepting the displayed deadline at face value, it identifies the function controlling the countdown from the DOM, sets the timer to five seconds, and waits for the apparent deadline to expire. It observes that the countdown resets while the promotion remains accessible, allowing it to establish that the timer creates artificial urgency rather than enforcing a genuine deadline. ScamFerret, which cannot interact with the page or examine its dynamic behavior, does not identify this deceptive mechanism.

During its investigation, ScamFerret also generates the search queries ``wego-france.com scam'' and ``YETI backpack giveaway'' to seek external evidence. The first leads it to a Trustpilot page for Wego.com, where it observes over 2,000 reviews and a rating of 3.1 stars. Despite the average rating, ScamFerret interprets the number of reviews as a possible indication of legitimacy, although it cannot establish a connection between the investigated domain and the official Wego website. The second query returns an official YETI giveaway page, \textsf{https://www.yeti.com/mystery-tundra-giveaway.html}, from which ScamFerret incorrectly infers that a backpack giveaway is taking place and that the investigated website might therefore have some legitimacy. Although it acknowledges that scammers could exploit such promotions, these search results introduce uncertainty rather than helping it establish whether the investigated website is malicious.

The most significant difference between the two investigations emerges when examining the reward-claiming process. Both agents identify the initial survey, but only ReDence proceeds to interact with it. Using \emph{Interact Page}, ReDence completes the four survey questions and retrieves the updated page after each action. When the survey reveals a gift-box game, it adapts its investigation to determine how the prize is awarded and what follows. The first selected box is empty and prompts another attempt, while the second produces a winning message, a prize identifier, and instructions leading to a shipping page. ReDence continues interacting with the website, entering dummy name and address information into the shipping form, which subsequently leads to a page requesting payment information. ScamFerret, in contrast, cannot answer the survey questions or interact with the gift boxes. Consequently, it does not uncover the game, winning message, or subsequent requests for shipping information and credit-card details.

These differences ultimately lead the two agents to conflicting conclusions. ReDence connects the impersonated retailer branding, fabricated review engagement, resetting countdown, and deceptive reward-claiming process to the collection of personal and payment information. Based on this accumulated evidence, it classifies the website as a \textbf{sweepstakes scam}. ScamFerret, meanwhile, bases its conclusion primarily on the mismatch between the investigated domain and Wego's official website, the positive Facebook-style reviews, and the presence of a survey. It states that these artifacts do not ``comprehensively'' establish maliciousness and recommends further investigation. For its binary classification, ScamFerret returns \textbf{False (Not malicious)} and assigns the category \textbf{``Online shopping.''} Thus, while both agents encounter several of the same initial indicators, ReDence's ability to visually interpret website content, interact with dynamic elements, and adapt its investigation allows it to uncover evidence that ScamFerret misses, ultimately resulting in a more complete and accurate assessment of the website.


\providecolor{ledgernavy}{HTML}{244A73}
\providecolor{numberblue}{HTML}{315B87}
\providecolor{purpletext}{HTML}{603B64}
\providecolor{reportgold}{HTML}{896018}

\tcbset{
    evidenceexcerpt/.style={
        enhanced,
        boxrule=0.5pt,
        fonttitle=\bfseries,
        fontupper=\small,
        toptitle=4pt,
        bottomtitle=2pt,
        sharp corners,
        boxsep=0pt,
        left=12pt,
        right=12pt,
        top=8pt,
        bottom=10pt,
        width=\linewidth,
        unbreakable
    }
}

\lstdefinestyle{scamferretjson}{
    basicstyle=\ttfamily\footnotesize,
    columns=fullflexible,
    keepspaces=true,
    breaklines=true,
    breakatwhitespace=true,
    showstringspaces=false,
    morestring=[b]",
    stringstyle=\color{purpletext},
    morekeywords={false,true},
    keywordstyle=\color{purpletext},
    aboveskip=0pt,
    belowskip=0pt,
    frame=none,
    numbers=none
}

\begin{figure}[t]
\centering
\setlength{\abovecaptionskip}{8pt}

\begin{tcolorbox}[
    evidenceexcerpt,
    colback=gray!3,
    colframe=blue!70!black,
    title={ReDence: Evidence Ledger Excerpt},
    colbacktitle=blue!70!black,
    coltitle=white
]

\textcolor{ledgernavy}{
    \normalsize\textbf{Evidence ID: E001}
}

\begin{itemize}[
    leftmargin=0.8em,
    label=\textcolor{numberblue}{\scriptsize$\bullet$},
    itemsep=1pt,
    topsep=2pt,
    parsep=0pt
]
    \item \textcolor{numberblue}{\textbf{Observation:}}
    \textcolor{purpletext}{Facebook-style comments claim
    successful deliveries and include photographs presented
    as showing received backpacks.}

    \item \textcolor{numberblue}{\textbf{Source:}}
    \textcolor{purpletext}{
        \nolinkurl{wego-france.com/backpack},
        comment section.
    }

    \item \textcolor{numberblue}{\textbf{Evidence type:}}
    \textcolor{purpletext}{Website content.}

    \item \textcolor{numberblue}{\textbf{Status:}}
    \textcolor{purpletext}{Direct observation.}

    \item \textcolor{numberblue}{\textbf{Relevance:}}
    \textcolor{purpletext}{Presents apparent customer
    endorsements of the giveaway.}
\end{itemize}

\smallskip
{\color{blue!35}\hrule}
\smallskip

\textcolor{ledgernavy}{
    \normalsize\textbf{Evidence ID: E002}
}

\begin{itemize}[
    leftmargin=0.8em,
    label=\textcolor{numberblue}{\scriptsize$\bullet$},
    itemsep=1pt,
    topsep=2pt,
    parsep=0pt
]
    \item \textcolor{numberblue}{\textbf{Observation:}}
    \textcolor{purpletext}{Commenter names are not clickable,
    and reaction buttons do not respond.}

    \item \textcolor{numberblue}{\textbf{Source:}}
    \textcolor{purpletext}{
        \nolinkurl{wego-france.com/backpack},
        comment controls.
    }

    \item \textcolor{numberblue}{\textbf{Action:}}
    \textcolor{purpletext}{Attempt to follow commenter names
    and activate reactions.}

    \item \textcolor{numberblue}{\textbf{Evidence type:}}
    \textcolor{purpletext}{Website interaction.}

    \item \textcolor{numberblue}{\textbf{Status:}}
    \textcolor{purpletext}{Direct observation.}

    \item \textcolor{numberblue}{\textbf{Relevance:}}
    \textcolor{purpletext}{The displayed engagement cannot
    be checked or reproduced through these controls.}
\end{itemize}

\smallskip
{\color{blue!35}\hrule}
\smallskip

\textcolor{ledgernavy}{
    \normalsize\textbf{Key Evidence}
}

\begin{itemize}[
    leftmargin=0.8em,
    label=\textcolor{numberblue}{\scriptsize$\bullet$},
    itemsep=1pt,
    topsep=2pt,
    parsep=0pt
]
    \item \textcolor{numberblue}{\textbf{Claim (inference):}}
    \textcolor{purpletext}{The endorsements may simulate
    social engagement to reassure visitors.}

    \item \textcolor{numberblue}{\textbf{Evidence IDs:}}
    \textcolor{purpletext}{\textbf{E001, E002}.}

    \item \textcolor{numberblue}{\textbf{Explanation:}}
    \textcolor{purpletext}{The delivery claims
    (\textbf{E001}) provide apparent social proof,
    while the nonfunctional profile and reaction controls
    (\textbf{E002}) weaken its credibility.}
\end{itemize}

\end{tcolorbox}

\caption{ReDence report excerpt connecting claimed customer
endorsements with observations from testing the comment controls.}
\label{fig:redence_report_excerpt}
\end{figure}

\begin{figure}[t]
\centering
\setlength{\abovecaptionskip}{8pt}

\begin{tcolorbox}[
    evidenceexcerpt,
    colback=yellow!4,
    colframe=reportgold,
    title={ScamFerret: Report Excerpt},
    colbacktitle=reportgold,
    coltitle=white
]

\begin{lstlisting}[style=scamferretjson]
{
  "scam": false,
  "website_type": "Online Shopping",
  "reason": "... The positive Facebook-style comments have no profile links, but they could represent reviews intentionally highlighted to promote the offer. ... The retrieved text and hyperlinks do not comprehensively establish that this website is malicious. ... further investigation is recommended."
}
\end{lstlisting}

\end{tcolorbox}

\caption{ScamFerret report excerpt presenting its classification,
website category, and narrative explanation.}
\label{fig:scamferret_report_excerpt}
\end{figure}

Moreover, we also find that ScamFerret’s report lacks the evidence-grounded structure of ReDence’s reports (retained from the Mentor model), as illustrated in Figures~\ref{fig:redence_report_excerpt} and~\ref{fig:scamferret_report_excerpt}. Beyond documenting individual findings, ReDence explains how they jointly support its interpretation. For example, E001 records the comments claiming successful deliveries, while E002 documents the nonclickable commenter names and unresponsive reaction controls. The key-evidence section references both entries, explaining that the delivery endorsements provide apparent social proof, while the nonfunctional controls weaken its credibility.  ScamFerret instead suggests that the comments could be reviews intentionally highlighted to promote the offer, without connecting this interpretation to documented interaction checks. Thus, not only does it find more features during its investigation, but can also present it in a way that allows an analyst to examine both the supporting observations and the reasoning that connects them, providing a clearer basis for evaluating whether the interpretation is justified.

\subsection{Reporting}
\label{redence-reporting}
We also evaluate whether Mentor’s reporting capabilities transfer to ReDence. Following the procedure described in Section~\ref{evaluating_reports}, we randomly select five malicious websites from each of the twelve scam categories, comprising the seven original SE scam categories and five held-out categories, yielding 60 malicious websites. We also include 60 benign websites. Our coder first investigates each website independently, then evaluates the reports generated by ReDence, Mentor, and ScamFerret using evidence coverage (EC) and auditable evidence coverage (AEC).

The coder correctly identifies 37 of the 60 malicious websites, returns inconclusive assessments for 13, and incorrectly classifies the remaining 10 as benign. We find that the coder correctly identifies all five credential-phishing websites,  but performance is less consistent across the other categories. This pattern further highlights the difficulty of recognizing scams within a limited investigation period when their deceptive behavior requires exploration beyond the landing page. Among the 60 benign websites, the coder correctly classifies 51, returns inconclusive assessments for 5, and incorrectly labels 4 as malicious. Even when the coder reaches the correct classification, the reports provide additional relevant evidence. Across the 37 websites correctly identified as malicious, the coder records an average of 2.8 distinct relevant findings per website, compared with 7.0 for ReDence, 6.5 for Mentor, and 3.4 for ScamFerret. ReDence therefore identifies slightly more relevant findings than Mentor and more than twice as many as ScamFerret on the same set of websites.

Over the sample, ReDence achieves a median Evidence coverage (EC) score of 0.87 and auditable evidence coverage (AEC) of 0.81. Mentor achieves corresponding scores of 0.84 and 0.77, closely matching its performance on the ground-truth dataset. On the other hand, ScamFerret obtains scores of 0.44 and 0.30 for EC and AEC scores respectively. This shows a modest advantage for ReDence over Mentor and a significantly larger advantage over ScamFerret in both capturing relevant findings and documenting sufficient supporting material for analysts to verify them.

\section{Conclusion}

Our findings demonstrate the benefits of ReDence's hybrid architecture, where the specialized Decider model accelerates decision-making and substantially reduces investigation time, while enabling ReDence to outperform both its larger Mentor model and state-of-the-art scam detectors. Furthermore, its agentic Investigator produces evidence-grounded reports that provide analysts with additional findings and contextual information they may have overlooked during manual investigations, supporting more informed and verifiable assessments of website legitimacy.


{\footnotesize \bibliographystyle{IEEEtran}
\bibliography{refs}}

%
\IEEEpeerreviewmaketitle

\end{document}